\documentclass[12pt]{article}
\usepackage{graphicx}

\usepackage{pslatex}
\usepackage{epstopdf}

\def\la{{\langle}}
\def\ra{{\rangle}}

\def\la{{\langle}}
\def\ra{{\rangle}}
\newcommand{\beq}{\begin{equation}}
\newcommand{\eeq}{\end{equation}}
\newcommand{\beqa}{\begin{eqnarray}}
\newcommand{\eeqa}{\end{eqnarray}}

\begin{document}
\baselineskip 14pt
\begin{center}
\vspace*{2cm}
{\Large\bf Feynman-path analysis of Hardy's paradox:
measurements and the uncertainty principle \vspace*{1cm}\\}

{\large D. Sokolovski}$^a$, {\large I. Puerto Gim\'enez}$^b$,
{\large R. Sala Mayato}$^b$ \footnote{E-mail address: 
dsokolovski@qub.ac.uk (D. Sokolovski); rsala@ull.es (R. Sala Mayato)}\\

$^a${\it School of Mathematics and Physics,\\
Queen's University of Belfast, Belfats, BT7 1NN, United Kingdom\\}

$^b${\it Departamento de F\'\i sica Fundamental II,\\
Universidad de La Laguna, La Laguna, 38204, S/C de Tenerife, Spain}

\end{center}

\pagestyle{plain}
\begin{center}
{\bf Abstract}
\end{center}
\begin{small}

Hardy's paradox is analysed within Feynman's formulation
of quantum mechanics. A transition amplitude is represented
as a sum over virtual paths which different intermediate measurements
convert into different  sets of real pathways.
Contradictory statements emerge when applying to the same statistical ensemble.
The ``strange'' weak values result is also investigated in this context.
%
%

\end{small}

\section*{1. Introduction}

The Hardy's paradox first introduced in \cite{HARDY}
continues to attract attention in the literature 
\cite{VHARDY, HSTAPP, UnHARDY, AHARDY, HANERT, LUND1, LUND2}.
The paradox consists in that, in a two-particle interferometer set up,
detection of particle(s) in different arms yields results obviously incompatible 
with each other.
One way to resolve the paradox is by noting that the conflicting results 
refer not to the same but to different physical situations thereby 
avoiding counterfactual reasoning - in the words of Ref. \cite{UnHARDY}
``talking about the values of non-measured attributes''.
The legitimacy of counterfactual statements was explored
in \cite{HSTAPP, UnHARDY} within the framework of formal
logic.
A different resolution involving inaccurate or weak quantum 
measurements was proposed in \cite{AHARDY}.
The purpose of this paper is to analyse the Hardy's paradox, counterfactual
statements and the weak measurements
in terms of virtual (Feynman) paths and the uncertainty principle \cite{Feyn1,Feyn2}.
In Feynman's quantum mechanics a transition probability
amplitude is found by adding amplitudes for all interfering paths
which, together, form a single indivisible pathway connecting 
the initial and final states of the system. The same paths can
be either interfering or exclusive alternatives, depending on whether
or not the system interacts with other (e.g. meter) degree's of freedom.
In addition, Feynman's  uncertainty principle states that any determination
of the path taken must destroy the interference between the 
alternatives \cite{Feyn1}.

The simplest illustration of the 
dangers of counterfactual reasoning is Young's double-slit 
experiment. One observes the probabilities with which 
electron starting in its initial state (source) reaches
a variety of final states (points on the screen) via single pathway
comprising both slits number $1$ and $2$ so that
an interference pattern is produced.
An accurate observation of the slit chosen by an electron
produces a system in which each final state can be reached 
via two real pathways (one through each slit) travelled with probabilities
$P_1$ and $P_2$. It is a {\it different} system: the interference
pattern is destroyed, and the probabilities to arrive in the final state
do not agree with the unobserved ones. Answering the ``which way?''
question by attributing the probabilities $P_1$ and $P_2$ to 
unobserved system would
constitute a counterfactual statement which is, obviously, wrong.
Below we will show that the Hardy's set up is equivalent to a three-slit
Young experiment, and that, in a similar way, the contradicting statements at the centre
of the ``paradox'' refer to different sets of real pathways produced by
different intermediate measurements and not to a single statistical ensemble.

The authors of \cite{AHARDY} have come to the defence 
of counterfactual reasoning suggesting to resolve the paradox
with the help of weak measurements \cite{Ah1, Ah2, ABOOK, 3B1}
performed
by a meter whose interaction with the system is so weak, that 
the interference between different paths is not destroyed.
There are two distinct issues associated with the weak measurements.
The answer to the first (and easier) question of whether they can be
performed in practice is yes \cite{HANERT, LUND1,LUND2,OPT}.
The second (and, to our knowledge not yet fully answered) question concerns the interpretation
of the weak results. Weak measurements are often seen ``as an extension
to the standard von Neumann model of measurements'' \cite{LUND2}
which allows one to refer to some of the outcomes as ``strange and surprising'' \cite{AHARDY}.
The problem is captured by the following simple example: consider applying a weak measurement
to determine the slit number in the above double-slit experiment.
For a particle arriving  near a minimum of the interference pattern one can obtain, say,
a number $10$ \cite{SWEAK} (a similar unusual value has been obtained
in the optical realisation of the experiment in \cite{OPT}).
This is not a valid (hence ``surprising'') slit number as there are only two holes 
drilled in the screen \cite{FOOTSLIT}.
Possible interpretations are that (i) the weak measurement reveals some new information about how
the  particle traverses the screen or, (ii) that in the weak limit the meter is not working properly.
Here we will follow Ref. \cite{SWEAK} in adopting the second 
view and turn to the uncertainty principle for an explanation. If the measured 
result is wrong, what is the correct answer to the ``which slit?'' question?
The uncertainty principle tells us that for an unobserved electron the
paths through the first and the second holes form (and this is the only paradox of quantum 
mechanics \cite{Feyn2}) a single indivisible pathway so that the correct answer does not exist. Thus a weak meter must either fail or the uncertainty principle 
would be proven wrong. The meter does fail by producing an answer which
appears to have nothing to do with the original question.
Mathematically, a weak value is an improper average obtained
with an alternating non-probabilistic amplitude distribution \cite{SWEAK} and as such
is not tied to its support (in this case, slit numbers $1$ and $2$), and serves mostly
to demonstrate that the particle cannot be seen as passing through a slit
with any particular probability.
We will argue further that the same can be said  about the ``resolution'' of the Hardy's paradox, 
which just as the above example relies on the surprising value $-1$ of the pair occupation number.

The rest of the paper is organised as follows:
in Section 2 we briefly describe measurements as a way of converting interfering virtual paths
into exclusive real ones.
In Section 3 we discuss the three-box case of Aharonov {\it et al} \cite{3B1} 
and its relation to the uncertainty principle.
In Section 4 we identify the virtual paths for the Hardy's set up.
In Section 5 we analyse real pathways produced from these by different measurements.
In Section 6 we consider differences between measurements
with and without post-selection. 
In Section 7 we discuss weak measurements for the Hardy's scheme and show
that a different choice of the final state can lead to extremely large ``anomalous''
weak occupation numbers.
Secttion 8 contains our conclusions.

\section*{2. Quantum measurements and virtual paths}

Consider a quantum system with a zero Hamiltonian
$\hat{H}=0$ in a $N$-dimesional Hilbert
space.  
The system is prepared (pre-selected) in some state
$|i\ra $ and, at a later time $T$, observed (post-selected) in
a final state $|f\ra $.
It is convenient to choose an othonormal basis $\{|n\ra\}, \quad n=1,2..N$
corresponding to the ``position'' operator \cite{posop}
$\hat{n} \equiv \sum_{n=1}^N |n\ra n \la n|$.
In general, the transition amplitude can be written as a 
sum over all virtual paths $n(t)$ which take the values
$1,2... N$ at any given time $t$ \cite{SR1,SR2}.
Since  $\hat{H}=0$, there are only $N$ constant paths 
$n(t)=1,2... N$ and the path decomposition  of the transition
amplitudes takes a simple form 
\begin{equation} \label{0.1}
\la f|i\ra = \sum_{n=1}^N\la f|n\ra \la n|i\ra \equiv \sum_{\{n\}}\Phi \{n\}\, ,
\end{equation}
where $\Phi \{n\}=\la f|n\ra \la n|i\ra$ is the amplitude for the $n$-th path.
With these notations the problem becomes equivalent
to a $N$-slit Young experiment with $N$ discrete final 
destinations (positions on the screen) given by the chosen
final state $|f\ra $ and any $N-1$ orthonormal states spanning the 
Hilbert subspace orthogonal to $|f\ra $.
The virtual paths provide a convenient way to describe an intermediate von Neumann
measurement at $0<t<T$ 
of any  operator $F(\hat{n})$ which commutes with the position $\hat{n}$,
\begin{equation} \label{0.2}
F(\hat{n})\equiv\sum_{n=1}^N |n\ra F(n) \la n|\, ,
\end{equation}
where $F(n)$ is an arbitrary function. For an accurate measurement
the initial pointer position $f$  is known exactly, its initial state is
the delta-function $\delta(f)$ and the probability amplitude $\Phi(f)$ to register a
meter reading $f$ is given by 
\begin{equation} \label{0.3}
\Phi(f)  = \sum_{\{n\}}\delta(f-F(n))\Phi \{n\}\, .
\end{equation}
It is readily seen that  if $F(\hat{n})$ has $M<N$ distinct eigenvalues
$F_1$, $F_2$,... $F_M$
with the multiplicities $m_1,m_2,...m_M$, respectively, 
 $N$ virtual paths are divided into $M$ exclusive classes.
Each such class can be seen to form a real indivisible pathway to which one
can assign a probability $P^{f \leftarrow i}_{\{m\}}$, with
which a reading $f=F_m$ would occur if the measurement
is performed. 

The transition probabilities are, in general, altered 
by the measurement,
\begin{equation} \label{0.6}
\tilde{P}^{f\leftarrow i}\equiv \sum_{j=1}^M|\sum_{n=1}^N \delta(F(n)-F_j) \Phi \{n\}|^2  
\ne |\la f|i\ra |^2\, .
\end{equation}
Note that if all eigenvalues of $F(\hat{n})$ are the same, no new real pathways are produced,
nothing is measured and no perturbation is incurred.
Note also that Eq. (\ref{0.6}) is just an example of the Feynman's rule for
assigning probabilities \cite{Feyn2} which
prescribes adding amplitudes for the interfering, 
and probabilities for the exclusive alternatives.

Another parameter which affects both the obtained 
information and the incurred perturbation is the accuracy 
of the measurement $\Delta f$.
If the meter's
sharp initial state is replaced with
a distributed one,  $\delta(f) \rightarrow G(f)$, where $G(f)$
can be chosen a Gaussian with a width $\Delta f$,
the amplitude to obtain a reading $f$ is given by 
\begin{equation} \label{0.7}
\Psi(f)  = \int df' G(f-f')\Phi(f')=\sum_{n}G(f-F(n))\Phi \{n\}\, .
\end{equation}
In general, an inaccurate measurement produces a continuum of real pathways
labelled by the variable $f$.
The same path $\{n\}$  contributes to many pathways,
its contribution to $\Psi(f)$ being $G(f-F(n))\Phi\{n\}$.
The perturbation decreases with the increase of the 
uncertainty $\Delta f$ and becomes negligible
for $\Delta f >> \delta f$ where $\delta f$ is the 
difference between the largest and the smallest eigenvalue
of $F(\hat{n})$. Application of the meter under these conditions
leads to so-called weak measurements first proposed in \cite{Ah1, Ah2, ABOOK}
and recently discussed in \cite{SWEAK}. We will return to weak measurements in Section 7.

\section*{3. The uncertainty principle and the three-box example}

The uncertainty principle states that  \cite{Feyn1}: 
``Any determination of the alternative taken by a process
capable of following more that one alternative
destroys the interference between alternatives''.
An illustration of what the principle may mean in practical
circumstances is
provided by the three-box
case by Aharonov {\it et al} \cite{ABOOK, 3B1, 3B2, 3box, SPR}.
Consider a three-state system with a zero Hamiltonian
pre- and post-selected in states such that the amplitudes in 
(\ref{0.3}) are ($\beta < 1$ is real)
\begin{equation} \label{1.1}
\Phi\{1\}=\beta , \quad \Phi \{2\}=-\beta \quad and \quad \Phi \{3\}=-\beta 
\end{equation}
so that for an unobserved system we have
\begin{equation} \label{1.2}
\la f|i\ra = \sum_{\{n\}}\Phi \{n\}=\beta\, .
\end{equation}
One sees that there is a cancellation between the paths but cannot decide
whether it is paths $\{1\}$ and $\{2\}$ or $\{1\}$ and $\{3\}$ that 
make each other redundant.
All three paths must therefore be treated as a single
indivisile pathway, denoted $\{1+2+3\}$,
travelled with the probability $P_{\{1+2+3\}}^{f\leftarrow i}=
|\Phi\{1\}+\Phi\{2\}+\Phi\{3\}|^2=\beta^2$.

If one decides to accurately measure the projector on $|2\ra$, whose matrix form is
\begin{equation} \label{1.2a}
\hat{P}_2\equiv diag(0,1,0)\, , 
\end{equation}
in order to see that the particle does indeed follow the path $\{2\}$,
the meter will read $1$ confirming the assumption and the 
particle will be detected in $|f\ra$ with the same probability
$P^{f\leftarrow i}=\beta^2$.
Perhaps surprisingly, measurement of the projector
\begin{equation} \label{1.2b}
\hat{P}_3\equiv diag(0,0,1)
\end{equation}
will confirm that particle always travels along the path $\{3\}$.
Assuming that these results refer to the same system (ensemble)
prompts a somewhat paradoxical conclusion that the particle is in several 
places simultaneously \cite{ABOOK,3B1}.
This ``paradox'' disappears \cite{SPR} once one notices
that measurement of $\hat{P}_2$ creates a new network
in which a final state can be reached via two real pathways, $\{2\}$, and a coherent superposition of the two remaing ones,
$\{1+3\}$. In a similar way, measureament of $\hat{P}_3$
fabricates yet another different statistical ensemble with two real pathways, $\{3\}$ and $\{1+2\}$.
What is true for one ensemble is not true for the other
even though both are produced by observations made on the
same system.
Equally, neither measurement reveals what ``actually''
happens in the unobserved system where the ``which way?'' information is lost, in accordance
with the uncertainty principle, to quantum interference.

\section*{4. Virtual paths in the Hardy's setup}

\begin{figure}
\vskip0.5cm
\includegraphics[width=12cm]{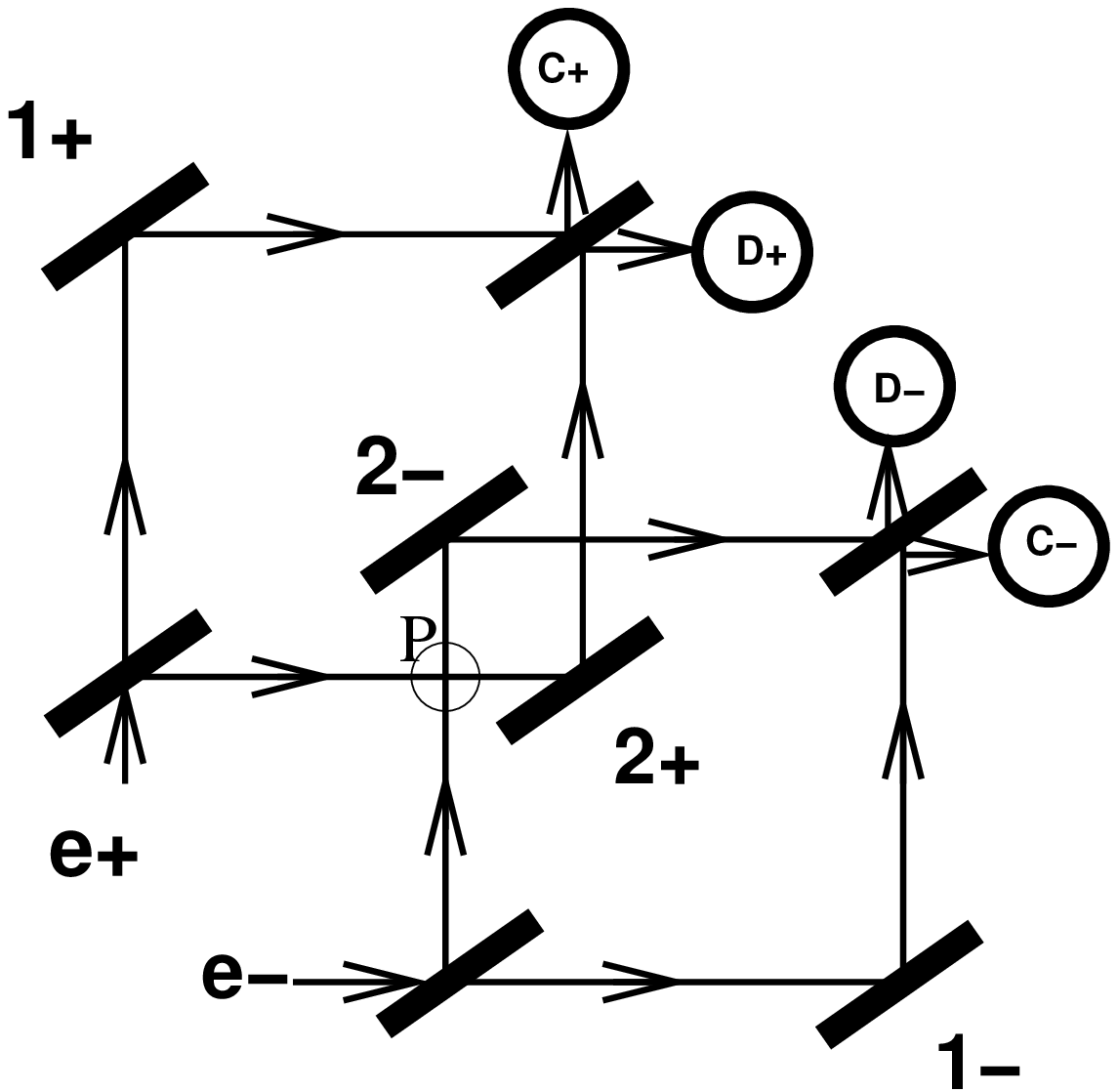}
\caption {The Hardy's set up of mirrors and beam splitters.
An electron ($e-$) and a positron ($e+$) are simultaneously
injected into the system. Their simultaneous presence in the 
overlapping arms $2-$ and $2+$, respectively, leads to certain
annihilation.}
\label{FIGF}
\end{figure}

\begin{figure}
\vskip0.5cm
\includegraphics[width=14cm]{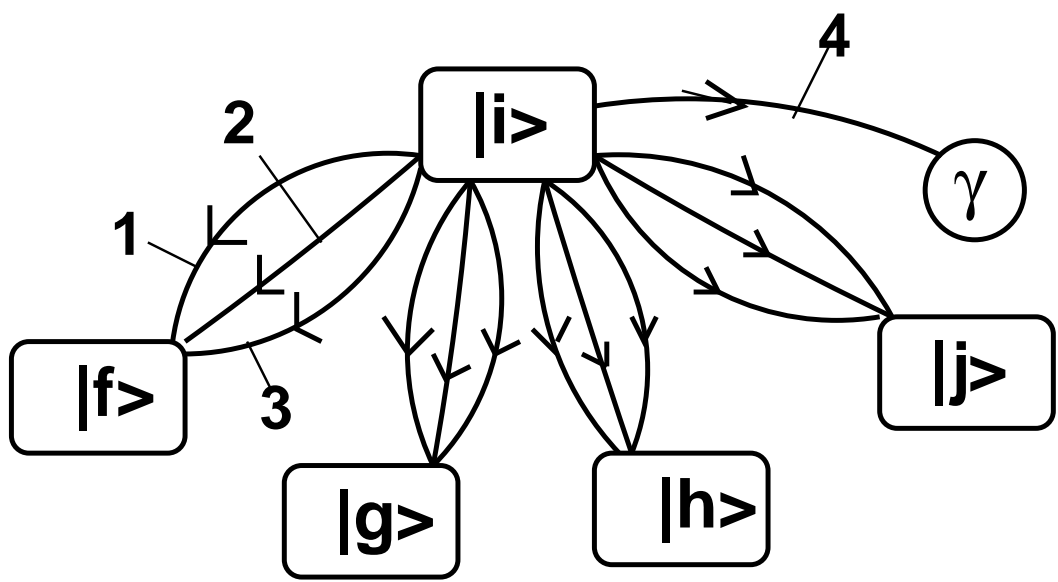}
\caption {Five final states and twelve paths connecting them 
with the initial state $|i\ra$.}
\label{FIG4b}
\end{figure}

The Hardy's set up shown in Fig. \ref{FIGF} has been discussed 
by many authors \cite{HARDY,VHARDY, AHARDY,HANERT,LUND1}
and a brief description will suffice here. An electron $(e-)$ and a positron
$(e+)$ are injected into their respective Mach-Zehnder  interferometers, each equipped with
two detectors, labelled $C-$, $D-$, $C+$ and $D+$.
So long as two interferometers are independent,
the outcome of the experiment consists in two detectors clicking
in coincidence. There are, therefore, four possible final outcomes.
Each particle has a choice of one of the two arms of the corresponding
interferometer, so that there are four virtual paths to reach each outcome.
The system is, therefore, equivalent to the four-slit Young diffraction 
experiment, with a minor distinction that there are only four discrete
final states (positions on the screen).
Hardy changes this arrangement by allowing two interferometer arms,
labelled $2-$ and $2+$ overlap so that if both particles are injected 
simultaneously and choose to travel along the
overlapping arms, annihilation follows with certainty.
Further, each interferometer is
tuned in such a way that neither
$D-$ nor $D+$ click if only an electron or a positron is injected.
Now the outcome of the experiment consist in two of the four detectors
clicking in coincidence or in a photon, $\gamma$, produced in annihilation.
There are, therefore, five possible final states.
Let the orthogonal state $|1-\ra (|2+\ra)$ and $|2-\ra (|1+\ra)$
correspond to the electron (positron) 
travelling via non-overlapping (overlapping) arms and vice versa
of the corresponding interferometer, respectively.
Similarly, the states $|1-\ra |1+\ra$ and $|2-\ra |2+\ra$
correspond to both particle travelling via non-overlapping and
overlapping arms.
The four paths can be labelled as follows:  
\begin{eqnarray} \label{3.0}
{\{1\}} & \quad {via} \quad & |1-\ra |1+\ra \nonumber
\\
{\{2\}} &\quad {via}  \quad & |1-\ra |2+\ra\nonumber
\\
{\{3\}} & \quad {via} \quad & |2-\ra |1+\ra \nonumber
\\  
{\{4\}} &\quad {via} \quad & |2-\ra |2+\ra\, .
\end{eqnarray}
Just past the point $P$ in Fig.1, the state
of the electron-positron pair is
\begin{equation} \label{3.1}
|i\ra =(|1+\ra |1-\ra
+|1+\ra |2-\ra
+|2+\ra |1-\ra
+|\gamma\ra)/2\, ,
\end{equation}
where $|\gamma \ra$ is the state of the photon produced in annihilation.
Of the five final states, the four corresponding 
to electron and positron arriving at the detectors $(D-,D+)$, $(C-,D+)$, $(D-,C+)$ and $(C-,C+)$,
respectively, are given by
\begin{eqnarray} \label{3.3f}
|f\ra =(|1-\ra-|2-\ra)(|1+\ra-|2+\ra)/2 
\end{eqnarray}
\begin{eqnarray} \label{3.3g}
|g\ra =(|1-\ra +|2-\ra)(|1+\ra-|2+\ra)/2 
\end{eqnarray}
\begin{eqnarray} \label{3.3h}
|h\ra =(|1-\ra-|2-\ra)(|1+\ra+|2+\ra)/2 
\end{eqnarray}
\begin{eqnarray} \label{3.3j}
|j\ra =(|1-\ra+|2-\ra)(|1+\ra+|2+\ra)/2\, .
\end{eqnarray}
Figure \ref{FIG4b} shows the network of paths connecting the initial and final states of the Hardy's 
system. 
There is only one path which connects the initial 
state with $|\gamma\ra$, while each of the final states
(\ref{3.3f}-\ref{3.3j})
can be reached via each of the pathways $\{1\}$, $\{2\}$
and $\{3\}$.
For example, the probability amplitude to reach a final
state $|z\ra$ via pathway $\{1\}$ is given by 
\begin{equation} \label{3.2}
\Psi^{z\leftarrow i}_{\{1\}} = \la z|1-\ra|1+\ra\la1+|\la1-|i\ra, \quad z=f,g,h,j,\gamma\, .
\end{equation}
Probability amplitudes for the paths $\{2\}$ and $\{3\}$
can be found in asimilar manner and are given in Table.1. 
Paths connecting different final states (see Fig.\ref{FIG4b}) are mutually exclusive and cannot 
interfere.  If no additional observations are made on the system, virtual paths ending in the same
final state are interfering alternatives and must be treated as a single real pathway.
Coherence between such paths can be destroyed, e.g., by an intermediate 
accurate measurement of an operator. As discussed in Section 2, each such measurement
would produce, depending on the multiplicity of the operator's  eigenvalues,
a number of additional real pathways.
In the next Section we will consider these measurements in more detail.

\section*{5. Accurate measurements and real pathways in Hardy's set up. Counterfactuals}

Following \cite{AHARDY} we wish to perform intermediate
measurements of ``pair occupation''  operators which establish
whether the electron and the positron are propagating along
specified arms of their respective interferometers. 
In the basis consisting of the states in the r.h.s. of Eq. (\ref{3.0})
these projectors take the form
\begin{eqnarray} \label{4.1}
\hat{ N}(1-|1+)=diag(1,0,0,0)
\end{eqnarray}
\begin{eqnarray} \label{4.2}
\hat{ N}(1-|2+)=diag(0,1,0,0)
\end{eqnarray}
\begin{eqnarray} \label{4.3}
\hat{ N}(2-|1+)=diag(0,0,1,0)
\end{eqnarray}
\begin{eqnarray} \label{4.4}
\hat{ N}(2-|2+)=diag(0,0,0,1)\, .
\end{eqnarray}
We will also require single particle occupation operators which 
establish whether the electron (positron) travels along the specified
arm, while the position of the other member of the pair remains
indeterminate,
\begin{eqnarray} \label{4.5}
\hat{ N}(1-)=\hat{ N}(1-|1+)+\hat{ N}(1-|2+) =diag(1,1,0,0)= 1-\hat{ N}(2-)
\end{eqnarray}
\begin{eqnarray} \label{4.6}
\hat{ N}(1+)=\hat{ N}(1-|1+)+\hat{ N}(2-|1+) =diag(1,0,1,0)= 1-\hat{ N}(2+)\, .
\end{eqnarray}
Consider now accurate measurements of these operators for 
the system post-selected in $|f\ra$ (electron and positron are detected
in $D-$ and $D+$, respectively). There are only three contributing paths, $\{1\}$, $\{2\}$
and $\{3\}$ with the amplitudes $1/4$, $-1/4$ and $-1/4$, so that the problem 
becomes equivalent to the three-box case of Section 3. If 
$\hat{ N}(1-|1+)$ is measured accurately, path $\{1\}$ becomes a real pathway, while the second
real pathway, $\{2+3\}$, is formed by interfering virtual paths $\{2\}$ and $\{3\}$ as
is shown in Fig. \ref{FIGA}.
The probabilities for these pathways are found by 
adding, where appropriate, the corresponding amplitudes in Table 1
and squaring the moduli,
\begin{eqnarray} \label{4.5a}
P^{f\leftarrow i}_{\{1\}}=|\Phi^{f\leftarrow i}_{\{1\}}|^2=1/16, \quad
P^{f\leftarrow i}_{\{2+3\}}=|\Phi^{f\leftarrow i}_{\{2\}}+\Phi^{f\leftarrow i}_{\{3\}}|^2=1/4\, .
\end{eqnarray}
Repeating the calculation for the case $\hat{ N}(1-|2+)$ is measured yields
\begin{eqnarray} \label{4.5b}
P^{f\leftarrow i}_{\{1\}}=|\Phi^{f\leftarrow i}_{\{2\}}|^2=1/16, \quad
P^{f\leftarrow i}_{\{1+3\}}=|\Phi^{f\leftarrow i}_{\{1\}}+\Phi^{f\leftarrow i}_{\{3\}}|^2=0\, ,
\end{eqnarray}
suggesting that if $D+$ and $D-$ click
(which happens with a probability of $1/16$)
{\it  (I) the electron and the positron always travel along the non-overlapping
and overlapping arms, respectively}. Similarly, for $\hat{ N}(2-|1+)$ one finds
\begin{eqnarray} \label{4.5c}
P^{f\leftarrow i}_{\{3\}}=1/16, \quad
P^{f\leftarrow i}_{\{1+2\}}=0\, ,
\end{eqnarray}
and might conclude, in contradiction to the above,
that {\it (II)  the electron and the positron always travel along the overlapping
and non-overlapping arms, respectively}.
For the single particle operators (\ref{4.5}-\ref{4.6}) real pathways and corresponding 
probabilities can be constructed in a similar manner and are given in Table 2.
Thus, measuring $\hat{ N}(1-)$  we find that {\it (III) the electron always travels
along the overlapping arm}. Measuring $\hat{ N}(1+)$ reveals that {\it (IV) the positron always travels
along the overlapping arm}. The italicised statements $(I)$,$(II)$,$(III)$ and $(IV)$,
if referred to the same system, would imply that  ``and electron and a positron in some way manage
to ``be'' and ``not be'' at the same time at the same location'' \cite{AHARDY}.
As in Section 3 the ``paradoxical''  nature of the Hardy's example is removed once one notices that
the above statements refer to different networks of classical (real) pathways produced from the
same parent unobserved system. That these networks are indeed different is 
seen already  from the fact that 
the transition probabilities to arrive in the final states $|g\ra$, $|h\ra$ and $|j\ra$ in Table 2
are different for each choice of the measured quantity. Note that for the final state $|f\ra$ the 
transition probability remains unchanged, but only due to the special choice
of the system's parameters.
Thus, only  the statement 
$(I)$ applies under the condition $\hat{ N}(1-|2+)$ is measured, while statements 
$(II)$, $(III)$ and $(IV)$
refer to unmeasured attributes and should be discarded.
The approach to resolving quantum interference ``paradoxes'' based on avoiding counterfactual
reasoning is by no means new \cite{HSTAPP, UnHARDY, AHARDY}. 
We note however, that the path analysis with its notion of converting interfering
virtual paths into exclusive real ones, provides a helpful insight into the argument.

\section*{6. ``Which way?'' probabilities without post-selection. The sum and the product 
rules}

In the above $P^{f\leftarrow i}_{\{1\}}=
|\Phi^{f\leftarrow i}_{\{1\}}|^2$  gave the probability 
for the pair to travel along the non-overlapping arms
provided the detectors $D+$ and $D-$ click in coincidence.
Alternatively, we may choose to record the frequency with
which the route is travelled regardless
of which detectors click, or just switch the 
detectors off completely.
Bearing in mind that the paths connecting different 
final states cannot interfere we find the corresponding
probability to be
\begin{eqnarray} \label{9.1}
P^{all\leftarrow i}_{\{1\}}=|\Phi^{f\leftarrow i}_{\{1\}}|^2
+|\Phi^{g\leftarrow i}_{\{1\}}|^2
+|\Phi^{h\leftarrow i}_{\{1\}}|^2
+|\Phi^{j\leftarrow i}_{\{1\}}|^2
=\la i|\hat{ N}(1-|1+)|i\ra\, .
\end{eqnarray}
The operator average in the r.h.s. of (\ref{9.1}) is the standard 
\cite{CT} expression for the probability of an outcome
for a system in the state $|i\ra$, 
which we have {\it derived} 
using (\ref{3.2}), completeness of the states
(\ref{3.3f})-(\ref{3.3j}) and the fact that
$\hat{N}(1-|1+)^2=\hat{N}(1-|1+)$.
Note that summation over all final outcomes 
has given $P^{all\leftarrow i}_{\{1\}}$  
properties not possessed by the state-to-state probabilities
$P^{z\leftarrow i}_{\{1\}}$, $z=f,g,h,j$.
For example, while $P^{z\leftarrow i}_{\{1\}}$ clearly
varies with the choice of the final state $|z\ra$,
$P^{all\leftarrow i}_{\{1\}}$ remains the same
for all choices of orthogonal final sates 
$|f\ra$, $|g\ra$, $|h\ra$ and $|j\ra$, and
only depends on the inital state $|i\ra$.     
Also, from Table 2 one notes that
(we apologise for the somewhat awkward sentence about to follow)
the state-to-state probability for the electron to travel
along the non-overlapping arm if 
$\hat{ N}(1-)$ is measured
does not equal the sum of probabilites
for the electron and positron to travel 
along non-overlapping arms if $\hat{ N}(1-|1+)$
is measured,
and that for the electron to travel along non-overlapping
and the positron along the overlapping arms, respectively,
provided we measure $\hat{ N}(1-|2+)$, e.g.,
\begin{eqnarray} \label{9.2}
P^{f\leftarrow i}_{\{1+2\}}=0 \ne P^{f\leftarrow i}_{\{1\}}+ 
P^{f\leftarrow i}_{\{2\}}=1/8\, ,
\end{eqnarray}
even though $\hat{ N}(1-)=\hat{ N}(1-|1+)+\hat{ N}(1-|2+)$.
This should not come as a surprise since, as discussed above,
measurements of $\hat{ N}(1-)$, $\hat{ N}(1-|1+)$ and
$\hat{ N}(1-|2+)$ create three different statistical
ensembles 
and Eq. (\ref{9.2}) is nothing more than an indication of this
fact.
An additional ``sum rule'' is obtained only if the individual
``which way?'' probabilities are summed over all final states
\begin{eqnarray} \label{9.3}
P^{all\leftarrow i}_{\{1+2\}}\equiv
\la i|\hat{ N}(1-)|i\ra=
\la i|\hat{ N}(1-|1+)+\hat{ N}(1-|2+)|i\ra
= P^{all\leftarrow i}_{\{1\}}+P^{all\leftarrow i}_{\{2\}}\, .
\end{eqnarray}

A closely related subject is the ``failure'' of the product
rule for post-selected systems \cite{VHARDY, ABOOK, BUB}.
As was shown in the previous Section, a measurement of $\hat{ N}(2-)$, conditioned on detectors
$D+$ and $D-$ clicking at the same time, shows that
the electron always travels along its overlapping
arm. Measuring $\hat{ N}(2+)$ shows that the same can be
said about the positron as well (see Table 2).
Table 1 shows that a measurement of $\hat{ N}(2-|2+)$ gives $P^{f\leftarrow i}_{\{4\}}=|\Phi^{f\leftarrow i}|^2=0$ and reveals, as it should,
that the two particles cannot travel the overlapping
paths simultaneously as they would annihilate and never
reach the detectors. Morover, from Eqs. (\ref{4.1}) to (\ref{4.6}) it is clear that
$\hat{ N}(2-|2+)=\hat{ N}(2-)\hat{ N}(2+)$.
Again, assuming that all three results refer to the same statistical
ensemble leads to a contradiction, as the joint probability
of two certain events must also equal one. Hence
inapplicability of the product rule gives another evidence
that with post-selection a different ensemble is produced with each choice
of the measured quantity, even though all three operators
commute \cite{SPR}. 
It is a simple matter to verify that without post-selection
conditions $\la i|\hat{A}|i\ra=1$ and $\la i|\hat{B}|i\ra=1$
would force $\la i|\hat{A}\hat{B}|i\ra=1$ for any two
commuting operators $\hat{A}$ and $\hat{B}$.

One of the purposes of this Section is to 
stress the fundamental nature of the Feynman's rule
for assigning probabilities \cite{Feyn2} 
which we have used first for 
constructing $P^{z\leftarrow i}_{\{1\}}$, $z=f,g,h,j$
and then in the derivation
of Eq. (\ref{9.1}).
On the other hand, someone who chooses the operator average in Eq. (\ref{9.1})
as a starting point for 
defining quantum probabilities might find 
generalisation to pre- and post-selected ensembles
more difficult, and the apparent break down of the sum
and the product 
rules an unexpected property of post-selection.

\section*{7. Weak measurements in Hardy's set up }

In Ref. \cite{AHARDY} the authors argued against discarding counterfactual statements
on the ground that they can be - to some extent- verified simultaneously provided
the accuracy of the measurements is so low that the system remains essentially unperturbed.
This can be achieved by making the meter state $G(f)$ in (\ref{0.7}) very broad,
\begin{equation} \label{5.1}
G(f) \rightarrow \alpha^{-1/4} G(f/\alpha), \quad \alpha \rightarrow \infty
\end{equation}
so that the mean meter reading $\la f \ra$ takes the form
\cite{ABOOK, SWEAK} ($G(f)=G(-f)$)
\begin{eqnarray} \label{5.2}
\la f \ra \equiv \int f |\Psi (f)|^2 df /\int  |\Psi (f)|^2 df  \approx Re \sum_{n=1}^N F(n)\Phi\{n\}/ \sum_{n=1}^N \Phi\{n\}
\equiv \bar{F}\, ,
\end{eqnarray}
where $ \bar{F}$ is the weak value of the operator $F(\hat{n})$.
Since there are no {\it apriori} restrictions on the phases of (in general complex valued) amplitudes
$\Phi \{n\}$, $ \bar{F}$ is an improper non-probabilistic average \cite{SWEAK}
with the obvious properties
\begin{equation} \label{5.3}
\bar{1}=1\, , 
\end{equation}
\begin{equation} \label{5.4}
\overline{F_1+F_2}= \bar{F_1}+\bar{F_2}\, , 
\end{equation}
and 
\begin{equation} \label{5.5}
\bar{F}= F(m) \quad if \quad  \Phi\{n\}=\Phi\{m\}\delta_{nm}\, , 
\end{equation}
i.e., in the absence of interference, when a single pathway connects 
initial and final states of the system.
 
In the three-box case of Section 3 one finds $\bar{P}_2=1$ for the projector $\hat{P}_2$,
just as it would be if an accurate measurement was conducted.
Similarly, for the projector $\hat{P}_3$ one has $\bar{P}_3=1$, which suggests that, since 
weak measurements do not perturb, the two expectation values, normally
observed in two different experiments, can be obtained simultaneously.

With the $|f\ra \leftarrow |i\ra$  transition in the Hardy's set up being equivalent
to the three-box case we therefore have: 
\begin{equation} \label{5.6}
\bar{N}(1-|2+) = \bar{N}(2-|1+) =1\, .
\end{equation}
This would have confirmed that two scenarios 
{\it (i) electron in the non-overlapping, positron the overlapping arms}
and 
{\it (ii) electron in the overlapping, positron the non-overlapping arms}
take place at the same time, had not the weak value of the third pair 
occupation operator $\bar{N}(1-|1+) = 1-\bar{N}(1-|2+)-\bar{N}(2-|1+)$ 
turned out to be $-1$ as required by Eqs. (\ref{5.3})
and (\ref{5.4}).  The authors of Ref. \cite{AHARDY} see in this
``remarkable way'' in which quantum mechanics solves the paradox
of having more than one electron-positron pair involved.

Here we adopt a different view.
Clearly, $\bar{N}(1-|1+)=-1$ is not a valid pair occupational number
for the non-overlapping arms of the interferometers, and as long
as it is an essential part of the reasoning, we rather doubt the whole ``resolution''
of the above paradox.
It is easy to see how this anomalous value arises.
Measurement of $\bar{N}(1-|1+)=diag(1,0,0,0)$ creates two
pathways, $\{1\}$ with the probability amplitude $1/4$, labelled 
$1$, and $\{2+3\}$ with the amplitude $-1/2$, labelled $0$ in Fig. 4.
In the weak limit (\ref{5.1}), the two pathways interfere and we 
have, in fact, a double-slit experiment, where we try to determine
the chosen slit without destroying the interference pattern on the
screen, which the uncertainty principle forbids \cite{Feyn1,Feyn2}.
A weak meter complies with the uncertainty principle by yielding a 
value which may lie anywhere on the real axis, and not just between
$0$ and $1$. The mathematical reason for this  is that the averaging in
the r.h.s. of Eq. (\ref{5.2}) is done with an improper alternating distribution \cite{SWEAK}, 
and the negative occupation number just manifests the failure to sensibly answer the ``which way?'' question.
 
This can be illustrated further by considering a more extreme case with the 
final state (\ref{3.3f}) replaced by (we assume that we 
can do that) 
\begin{equation} \label{5.7}
|f\ra = (|1-\ra|1+\ra-|1-\ra|2+\ra-\epsilon |2-\ra|1-\ra
+\epsilon |2-\ra|2+\ra)/2^{1/2}(1+\epsilon)^{1/2}\, ,
\end{equation}
where $\epsilon$ is a parameter, so that for $\epsilon =1$ we recover
the original transition in the Hardy's set up. Note that 
$ \hat{N}(2-|1+)$ creates a single pathway $\{3\}$, since
the amplitude for $\{1+2\}$ vanishes, whereas for $\hat{N}(1-|1+)$
and $\hat{N}(1-|2+)$ there exist two pathways with non-zero
amplitudes. The weak pair occupation numbers in (\ref{5.2})  now are 
\begin{eqnarray} \label{5.8}
\bar{N}(1-|1+)&=&-1/\epsilon\nonumber
\\ 
\bar{N}(1-|2+)&=&1/\epsilon\nonumber
\\
\bar{N}(2-|1+)&=&1\, .
\end{eqnarray}
Similarly, for the single particle 
occupation numbers we find
\begin{eqnarray} \label{5.9}
\bar{N}(1+)&=&1-1/\epsilon\nonumber
\\ 
\bar{N}(2-)&=&1.
\end{eqnarray}
Now let $\epsilon \rightarrow 0$
so that the transition becomes very improbable. 
For $\epsilon = 10^{-6}$ the last of Eqs. (\ref{5.8}) suggests
that, in defiance of the uncertainty principle, we have established
that in a system not perturbed by observations
electron and positron always travel along the 
overlapping and non-overlapping arms, respectively.
One must, however, add that there are also a million electron-positron pairs
travelling  along the non-overlapping and overlapping arms, 
and also minus one million pairs in the non-overlapping arms.
We might try to retain only the results where a weak value is obtained under
the non-interference condition (\ref{5.5}) but then we would have to discard also
the value  $\bar{N}(1-|1+)=-1$ needed to balance the books in the original example
of Ref. \cite{AHARDY}.
A more realistic view \cite{SWEAK}  is that the unusual properties of the weak 
values signal a failure of our measurement procedure in the case
when it is expected to fail. This failure occurs in a consistent way and can easily be 
observed, e.g., in an optical experiment \cite{OPT}.
What one observes, however, are the properties of a meter 
in a regime where it ceases to be a proper meter,
not to be confused with the
attributes of the measured system which, quantum mechanics
tells us, simply do not exist. 
For example, treating $\bar{N}(1+)$
as the actual charge in the non-overlapping positron arm \cite{AHARDY} would 
mean that with no effort whatsoever (although not very often)
we may create an arbitrary large charge, perhaps exceeding the largest one allowed by the
relativistic quantum mechanics \cite{LAND}.

\section*{8. Conclusions and discussion}

In summary, Feynman path approach offers a convenient language
for describing some controversial aspects of quantum measurement theory.
A quantum system can be seen as arriving in a given final 
state via a set of virtual (Feynman) paths defined for a particular
``position operator'' $\hat{n}$, all of whose eigenvalues are non-degenerate.
With no other measurements performed, the paths form one indivisible (real) 
pathway. Intermediate von Neumann
measurements of operators which commute with $\hat{n}$ produce,
by combining Feynman paths into classes and destroying
coherence between them,  a network of real
pathways connecting the initial and final states of the system.
The probabilities which can now be assigned
to the pathways define a classical statistical ensemble which
one observes in an experiment. 
The following statement can be regarded as a corollary to  
the Feynman's uncertainty principle \cite{Feyn1}: 
each set of intermediate 
measurements produces a different network of real pathways, in such a way that
in general 
no properties of a network $A$ can be inferred from a network $B$, and,
similarly, no properties of the unobserved system can be inferred from either
of the networks.
It is readily seen that an attempt to ascribe properties of different networks
(e.g., the four italicised statements of Section 5) to a single (e.g., unobserved)
system may lead to contradiction. Such attempts constitute counterfactual reasoning
which, as has been noticed before \cite{HSTAPP, UnHARDY} ought to be avoided. 
For the Hardy's set up 
we find, for example, that 

$(I)$ electron and positron always travel along the non-overlapping and overlapping
arms, respectively, (path $\{2\}$ in (\ref{3.0}))
{\it if  $\{2\}$ is decohered, while coherence between Feynman paths $\{1\}$ and $\{3\}$ is not destroyed},

$(II)$ electron and positron always travel along the overlapping and non-overlapping
arms, respectively, (path $\{3\}$ in (\ref{3.0}))
{\it if $\{3\}$ is decohered, while  coherence between Feynman paths $\{1\}$ and $\{2\}$ is not destroyed},
\newline
which presents no contradiction with the qualifications added.

We argue further that weak measurements do not provide a 
sufficient justification for counterfactual reasoning.
It is important to go through the statements made in \cite{AHARDY}
in greater detail. 
It is true that where only one path contributes to a transition, a
weak value would coincide with that obtained in an ideal strong
measurement - Eq. (\ref{0.7}) shows that that is the case for a meter of 
arbitrary accuracy.
One cannot, however, consistently avoid the ``strange'' 
anomalous weak values where there is more than one interfering
paths (cf. the exceptionally large occupation numbers discussed
in Section 6). Reference  \cite{AHARDY} tells us that ``strangeness is not a
problem; consistency is the real issue''.
This is not quite so: one is lead to believe that a weak measurement is, in some sense,
a valid extension of the accurate von Neumann measurement.
This notion largely accounts for one's interest in the subject as well as
for  one's surprise when a ``strange'' reading, such as a slit number $10$ in the two-slit case,  is produced.
An analogy with a purely classical meter employed to measure
the slit number for a classical particle is helpful:
the meter malfunctions and produces readings of, say, $10$ if the slit
$1$ is used and $15$ if the slit $2$ is used.
The link between a reading and the measured attribute
of the system (slit number $1$ or $2$) is lost and,
unless one knows how to re-calibrate the meter, 
the measurement becomes a study of the meter's rather than
the system's properties.
The peculiarity of the quantum situation is that
a correct suitable answer to the question just does not exist under the weak 
conditions and a weak meter cannot be re-calibrated.
In consequence, a weak measurement ceases to be a valid extension of the von Neumann procedure,
and its result becomes a manifestation of the failure to find a suitable value of the 
system's intrinsic attribute which, quantum mechanics tells us,
is unavailable.
Accordingly, the meter, which can no longer be
a good one, behaves {\it as if} the slit number were $10$ or {\it as if}
there were $10^6$ electron-positron pairs in a situation where it can be verified independently
that only two holes have been drilled in the screen and only two particles were injected into the system.
Many authors  \cite{AHARDY, ABOOK, OPT} 
have emphasised the ``surprising'' aspect of such anomalous weak values.
It would, however, been far more surprising had weak measurements consistently produced
``suitable'' results consistent, for example, with unobserved particle passing through
each slit with a certain probability.
Existence of strange anomalous values is a proof of the contrary  and
can be seen as a direct consequence of the uncertainty principle \cite{SWEAK}.
That not all weak values are strange and that the strange ones usually occur where
there is destructive interference between virtual paths does not alter this conclusion.
Finally, it is not surprising that a weak meter and the weak values obey their own self-consistent
logic \cite{AHARDY}. This would be the case for any degree of freedom whose interaction with 
another system is described by a reasonable coupling.
This logic does not, however, extend to (non-existing) intrinsic properties 
of the measured system, as is required if one wishes to make an argument in favor of counterfactual reasoning.
For this reason, weak measurements do not extend the limit set by the uncertainty principle on what can be learned
about a quantum system. It seems appropriate to conclude with Feynman's own warning
against such an extension \cite{Feyn3}:
``Do not keep saying to yourself, if you can possibly avoid it, ``But how can it be like that?'' 
because you will get ``down the drain'', into a blind alley from which nobody has yet escaped.''

\section*{Acknowledgemnts}

R. Sala Mayato is greatful to acknowledge Ministerio de Educaci\'on y Ciencia,
Plan Nacional under grants No. FIS2004-05687 and No. FIS2007-64018, 
and Consejer\'\i a de Educaci\'on, Cultura
y Deporte, Gobierno de Canarias under grant PI2004-025. 
I. Puerto is greatful to acknowledge Ministerio de Educaci\'on y Ciencia under grant
AP-2004-0143.
D. Sokolovski is greatful to acknowledge Universidad de La Laguna for partial support under
project ``Ayudas para la estancia de profesores e investigadores no vinculados a la Universidad de La Laguna''.

\newpage
\vspace*{2cm}

\newpage

\newpage
\begin{figure}
\vskip0.5cm
\includegraphics[width=14cm]{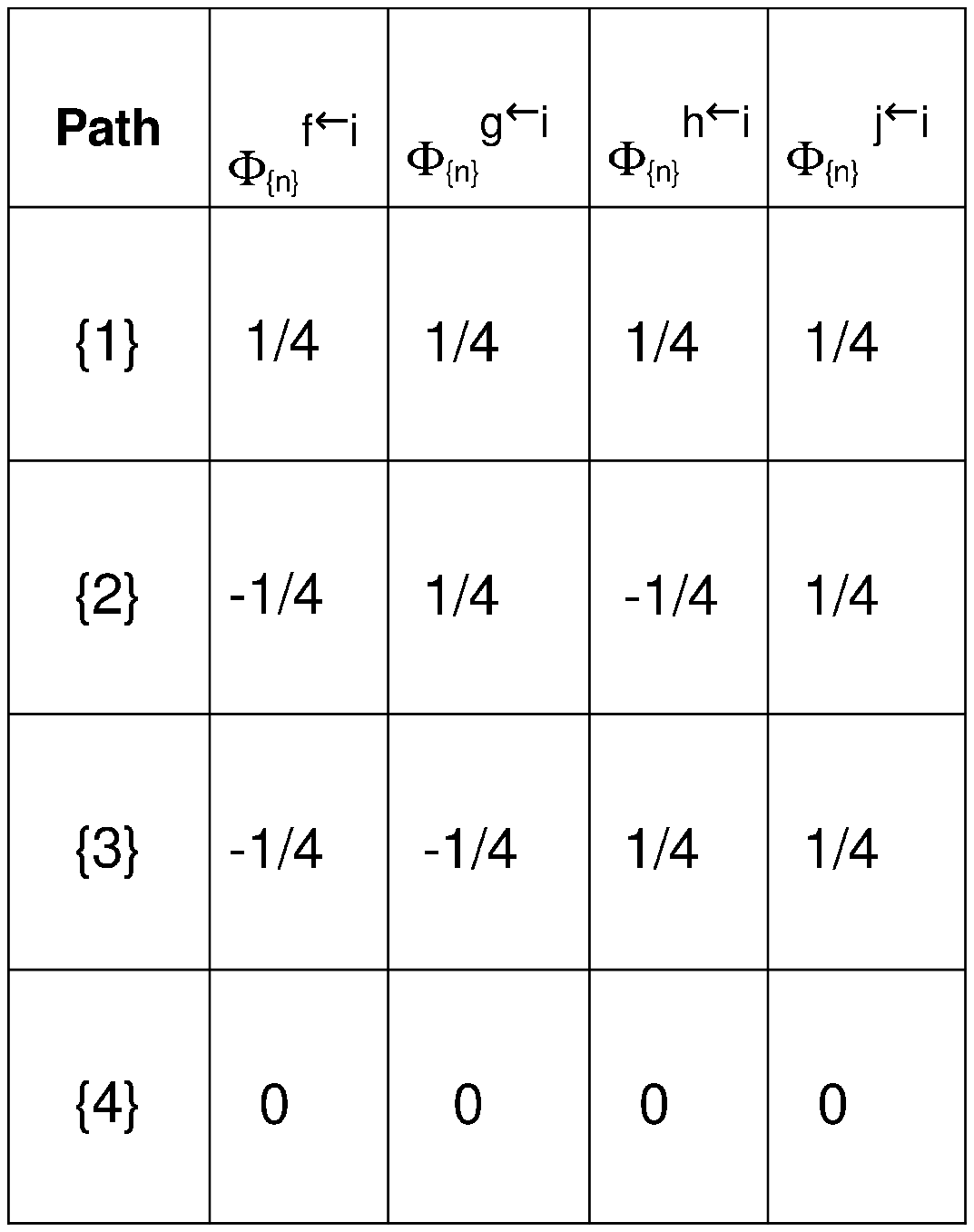}
\caption {Table 1. Probability amplitudes for the virtual paths
shown in Fig. \ref{FIG4b}.}
\label{fFIG3b}
\end{figure}

\newpage
\begin{figure}
\vskip0.5cm
\includegraphics[width=15cm]{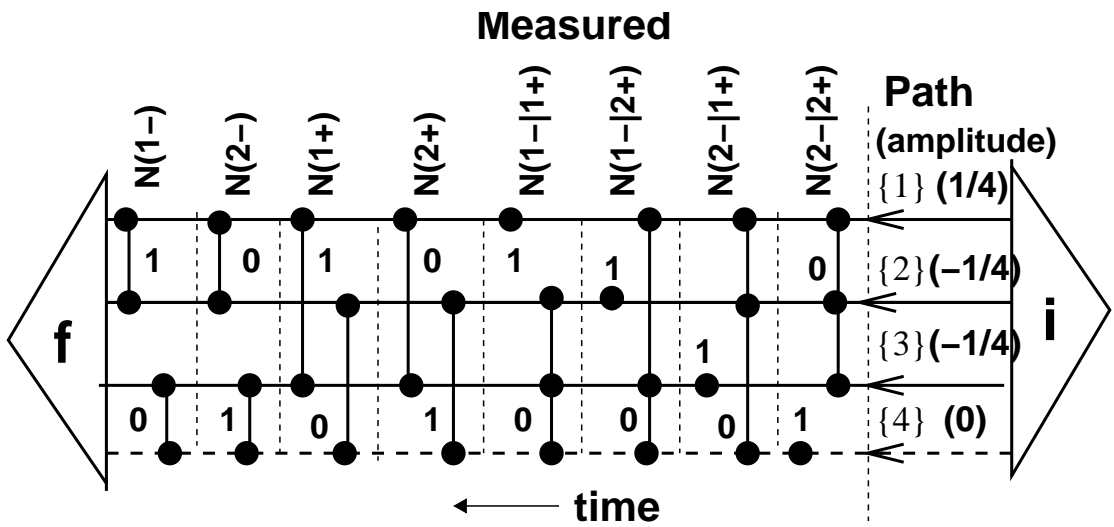}
\caption {Virtual paths in (\ref{3.0}) which connect the states $|i\ra$ and $|f\ra$.
The path $\{4\}$ has a zero amplitude due to annihilation, and is shown
by a dashed line. An accurate measurement of one of the operators
listed above creates two real pathways, each comprising the 
paths joined by the vertical lines. The nunmbers $1$ or $0$
next to the lines are the eigenvalues of the measured operator.}
\label{FIGA}
\end{figure}

\newpage
\begin{figure}
\vskip0.5cm
\includegraphics[width=16cm]{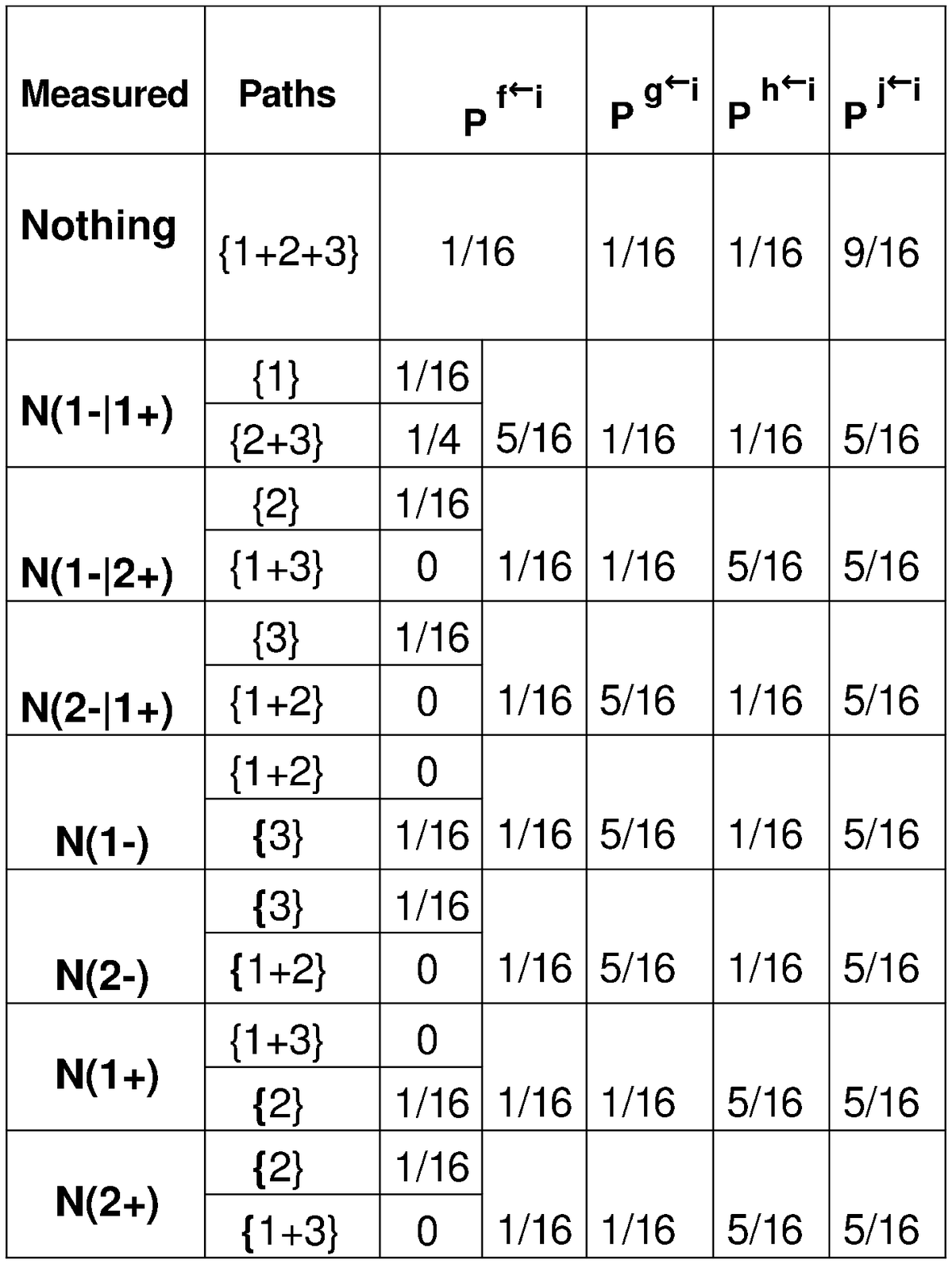}
\caption {Table 2. Transition probabilities for the final states in Fig.\ref{FIG4b}
if intermediate measurements in Fig.\ref{FIGA} are performed.
Also shown are the probabilities for the real pathways created by measurements
for the final state $|f\ra$ in (\ref{3.3f}). In all cases the annihilation occurs 
with probability of $1/4$.}
\end{figure}


\begin{thebibliography}{8.}
\addcontentsline{toc}{section}{References}

\bibitem{HARDY} L. Hardy, Phys. Rev. Lett. \textbf{68} (1992) 2981.

\bibitem{VHARDY} L. Vaidman, Phys. Rev. Lett. \textbf{70} (1993) 3369.

\bibitem{HSTAPP} H. S. Stapp
Am. J. Phys.  \textbf{65} (1997) 300; Am. J. Phys. \textbf{66} (1998) 924.

\bibitem{UnHARDY} W. Unruh, Phys. Rev. A \textbf{59} (1997) 126. 

\bibitem{AHARDY} Y. Aharonov, A. Botero, S. Popescu, B. Reznik and J. Tollaksen,
Phys. Lett. A \textbf{301} (2002) 130.

\bibitem{HANERT} S. E. Ahnert and M. C. Payne, Phys. Rev. A \textbf{70} (2004) 042102.  

\bibitem{LUND1} J. S. Lundeen and K. J. Resch,  Phys. Lett. A \textbf{334} (2005) 337.

\bibitem{LUND2} J. S. Lundeen, K. J. Resch and A. M. Steinberg, Phys. Rev. A \textbf{72} (2005) 016101.

 
\bibitem{Feyn1} R. P. Feynman and A. R. Hibbs,
\emph{Quantum Mechanics and Path Integrals}, (McGraw-Hill, New York 1965).

\bibitem{Feyn2} R. P. Feynman, R. B. Leighton and M. Sands, 
\emph{The Feynman lectures on physics: quantum mechanics},
(Addison-Wesley, 1965).


\bibitem{Ah1} Y. Aharonov, D. Z. Albert and L. Vaidman,
Phys. Rev. Lett. \textbf{60} (1988) 1351.

\bibitem{Ah2} Y. Aharonov and L. Vaidman, Phys. Rev. A \textbf{41} (1990) 11.

\bibitem{ABOOK}  Y. Aharonov and L. Vaidman, in {\it Time in Quantum Mechanics}, edited by
J. G. Muga, R. Sala Mayato and I. L. Egusquiza (Springer, 2002), pp. 369-413.

\bibitem{3B1} Y. Aharonov and L. Vaidman, J. Phys. A  \textbf{24} (1991) 2315.

\bibitem{OPT} N. W. M. Ritchie, J. G. Story and R. G. Hulet, Phys. Rev. Lett.
\textbf{66} (1991) 1107.

\bibitem{SWEAK} D. Sokolovski, Phys. Rev. A \textbf{76} (2007) 042125.


\bibitem{FOOTSLIT} 
The original paper on weak measureaments was \cite{Ah1} entitled
``How can a measurement of a spin $1/2$ give a result $100$?''.
While conceptually identical, the double-slit (Mach-Zehnder interferometer)
example provides somewhat stronger case against over-interpretation
of weak values. Unlike the value of a spin component, the number
of slits (arms) can established independently by purely classical methods. 
Thus any measurement result which appears to imply the existence
of more than two slits or arms would require an explanation of the 
apparent contradiction.

\bibitem{posop} Compare with the position operator for a particle in one 
spacial dimension, $\hat{x}=\int |x\ra x\la x| dx$.


\bibitem{SR1} D. Sokolovski and R. Sala Mayato, Phys.Rev. A \textbf{71} (2005) 042101.
\bibitem{SR2} D. Sokolovski and R. Sala Mayato, Phys.Rev. A \textbf{73} (2006) 052115;
\textbf{76} (2007) 039903(E).


\bibitem{3B2} T. Ravon and L. Vaidman, J. Phys. A  \textbf{40} (2007) 2873.

\bibitem{3box} C. R. Leavens, I. Puerto Gim\'enez, D. Alonso and R. Sala Mayato,
Phys. Lett. A \textbf{359} (2006) 416.


\bibitem{SPR} D. Sokolovski, I. Puerto Gim\'enez and R. Sala Mayato (unpublished).



\bibitem{CT} C. Cohen-Tanoudji, B. Diu and F. Lalo\"e,
\emph{Quantum Mechanics, Volume one}, 
(Hermann and John Wiley $\&$ Sons., 1977)

\bibitem{BUB} J. Bub and H. Brown, Phys. Rev. Lett. \textbf{56} (1986) 2337.

\bibitem{LAND} E. M. Lifshitz, V. B. Berestetskii and L. P. Pitaevski, \emph{Quantum 
Electrodynamics}, (Reed Elsevier, 1982)


\bibitem{Feyn3} R. P. Feynman, \emph{The character of physical law}, 
(Modern Library, 1994).

\end{thebibliography}
\end{document}